\documentclass[showpacs,aps,prb,reprint,superscriptaddress,preprintnumbers,nolinenumbers]{revtex4-1}

\usepackage{amsmath}
\usepackage{amssymb}
\usepackage{graphicx}
\usepackage{dcolumn}
\usepackage[dvipsnames]{xcolor}
\usepackage{amsthm,amssymb}
\usepackage{mathrsfs}
\usepackage{color}
\usepackage{endnotes}
\usepackage{bm}
\usepackage{epsfig}
\usepackage{array}
\usepackage{ulem}
\usepackage{lineno}
\newcommand{\stkout}[1]{\textcolor{red}{\ifmmode\text{\sout{\ensuremath{#1}}}\else\sout{#1}\fi}}

\newcommand{\etal}{{\it et al.}}

\newcommand{\CsVSb}{{CsV$_{3}$Sb$_{5}$}}
\newcommand{\RbVSb}{{RbV$_{3}$Sb$_{5}$}}
\newcommand{\KVSb}{{KV$_{3}$Sb$_{5}$}}
\newcommand{\AVSb}{{AV$_{3}$Sb$_{5}$}} 
\newcommand{\parallelsum}{\mathbin{\!/\mkern-5mu/\!}}

\begin{document}


\title{Anomalous Hall effect and two-dimensional Fermi surfaces\\ in the charge-density-wave state of kagome metal RbV$_{3}$Sb$_{5}$}

\author{Lingfei~Wang$^\S$}
\author{Wei~Zhang$^\S$}
\author{Zheyu~Wang$^\S$}
\author{Tsz~Fung~Poon}
\author{Wenyan~Wang}
\author{Chun~Wai~Tsang}
\author{Jianyu Xie}
\affiliation{Department of Physics, The Chinese University of Hong Kong, Shatin, Hong Kong, China}
\author{Xuefeng~Zhou}
\author{Yusheng~Zhao}
\author{Shanmin~Wang}
\affiliation{Department of Physics, Southern University of Science and Technology, Shenzhen, Guangdong, China}
\author{Kwing~To~Lai}
\email[]{ktlai@phy.cuhk.edu.hk}
\affiliation{Department of Physics, The Chinese University of Hong Kong, Shatin, Hong Kong, China} 
\affiliation{Shenzhen Research Institute, The Chinese University of Hong Kong, Shatin, Hong Kong, China}
\author{Swee~K.~Goh}
\email[]{skgoh@cuhk.edu.hk}
\affiliation{Department of Physics, The Chinese University of Hong Kong, Shatin, Hong Kong, China}



\date{\today}

\begin{abstract}
\AVSb\ (A=Cs, K, Rb) are recently discovered superconducting systems ($T_{\rm c}\sim0.9 - 2.5$~K) in which the vanadium atoms adopt the kagome structure. Intriguingly, these systems enter a charge-density-wave (CDW) phase ($T_{\rm CDW}\sim80-100$~K), and further evidence shows that the time-reversal symmetry is broken in the CDW phase. Concurrently, the anomalous Hall effect has been observed in \KVSb\ and \CsVSb\ inside the novel CDW phase.
Here, we report a comprehensive study of a high-quality RbV$_3$Sb$_5$ single crystal with magnetotransport measurements. Our data demonstrate the emergence of anomalous Hall effect in \RbVSb\ when the charge-density-wave state develops. The 
magnitude of anomalous Hall resistivity at the low temperature limit is comparable to the reported values in KV$_3$Sb$_5$ and CsV$_3$Sb$_5$. The magnetoresistance channel further reveals a rich spectrum of quantum oscillation frequencies, many of which have not been reported before. In particular, a large quantum oscillation frequency (2235 T), which occupies $\sim$56\% of the Brillouin zone area, has been recorded. For the quantum oscillation frequencies with sufficient signal-to-noise ratio, we further perform field-angle dependent measurements and our data indicate two-dimensional Fermi surfaces in RbV$_3$Sb$_5$.
Our results provide indispensable information for understanding the anomalous Hall effect and band structure in kagome metals \AVSb.

\end{abstract}

\maketitle

\section{Introduction} 
Kagome lattice intrinsically hosts the flat electronic band, van Hove singularities and the Dirac band~\cite{Kiesel2012,Kiesel2013,Wang2013}. Thus, when the charge degree of freedom is enabled, a kagome system provides an important platform to explore both the effect of electronic correlation and topological physics, and a rich variety of phenomena can be expected.
The recently discovered kagome metals AV$_3$Sb$_5$ (A=K, Rb, Cs), in which the V atoms form a perfect kagome net, are an ideal manifestation of the diverse phenomena this structural class can offer~\cite{Ortiz2019,Ortiz2020,Du2021,Chen2021a,Li2021,Liang2021,Zhao2021,Liu2021,Hu2021,Uykur2021,Zhou2021,Jiang2021,Kang2022,Kang2022a,Wu2022,Lou2022,Yu2021b,Yang2020,Yu2021,Wulferding2022,Jiang2022,Nie2022}. Indeed, calculations and experiments have revealed several Dirac-like band crossings near the Fermi energy with a non-zero $Z_2$ topological invariant~\cite{Ortiz2020,Fu2021,Liu2021}. Furthermore, these systems exhibit interesting and intertwining electronic phases, including a charge-density-wave (CDW) phase, electronic nematicity as well as superconductivity~\cite{Ortiz2019,Ortiz2020,Du2021,Chen2021a,Li2021,Liang2021,Zhao2021,Liu2021,Hu2021,Uykur2021,Zhou2021,Jiang2021,Kang2022,Kang2022a,Wu2022,Lou2022,Yu2021b,Yang2020,Yu2021,Liu2021,Wulferding2022,Jiang2022,Nie2022}. 

The CDW phase of \AVSb, which sets in at $T_{\rm CDW}\sim80-100$~K, has been argued to have an unconventional origin. Chiral charge order, the absence of acoustic phonon anomaly, and time-reversal symmetry breaking (TRSB) have been observed in the CDW phase~\cite{Feng2021,Li2021,Yu2021b,Yu2021c,Hu2022,Mielke2022,Khasanov2022,Guguchia2022,Jiang2021,Kenney2021,Ortiz2021b,Luo2022}. Below $T_{\rm CDW}$, a giant anomalous Hall effect (AHE) has been reported in \KVSb\ and \CsVSb, supporting the notion of TRSB~\cite{Yang2020,Yu2021b}. Given that the superconductivity, which sets in at $T_c\sim0.9-2.5$~K, arises from this unconventional metallic state, understanding the CDW phase and the phenomena associated with the CDW phase is essential for deepening the understanding of the superconducting state. Further evidence of the nontrivial interplay between the CDW state and superconductivity comes from pressure studies - while $T_{\rm CDW}$ is suppressed monotonically by pressure, $T_c$ shows an unusual double dome dependence on pressure~\cite{Yu2021,Chen2021a,Wang2021a,Du2021}.

Although all three \AVSb\ compounds show similar physical properties, \CsVSb\ is currently the most heavily studied, probably because it has the highest $T_c$. However, both \KVSb\ and \RbVSb\ should also be investigated to build an overall picture for understanding this family of V-based kagome superconductors. For instance, a $\mu$SR experiment shows that the superconducting energy gap can be tuned from nodal at ambient pressure to nodeless under pressure in \RbVSb\ and \KVSb~\cite{Guguchia2022}, but this tunability is absent in \CsVSb\ where the gap is found to be nodeless over a large pressure range~\cite{Gupta2022a}. Whether the observed nodal gap is related to TRSB needs to be addressed. However, the anomalous Hall effect (AHE), a signature of TRSB, has not been comprehensively reported in \RbVSb, to the best of our knowledge.  

Detailed studies of the Fermi surface via quantum oscillations have also been lacking in \RbVSb, again in stark contrast to the progress made in \CsVSb~\cite{Yu2021,Yu2021b,Fu2021,Ortiz2021,Gan2021,Zhang2022a,Chen2022,Huang2022,Shrestha2022, Chapai2022,Broyles2022}. In the pioneering work by Yin \etal~\cite{Yin2021}, only two quantum oscillation frequencies at a single field angle are reported. In particular, these frequencies are rather small -- $F_{\alpha}=34$~T and $F_{\beta}=117$~T, indicating the detection of small Fermi surface sheets. However, much larger frequencies up to 9930~T when the magnetic field is parallel to the $c$-axis have been reported in \CsVSb~\cite{Chapai2022}. Therefore, it is urgently needed to re-investigate the Fermi surface of \RbVSb\ via quantum oscillations. 

In this manuscript, we focus on \RbVSb~and conduct a series of magnetotransport experiments on high-quality single crystals. We first patch the gap in the literature by reporting the observation of a large AHE in \RbVSb. Next, we analyze the magnetoresistance, uncovering quantum oscillations (the Shubnikov-de Haas (SdH) effect) with 12 frequencies. Notably, the largest frequency detected is 2235~T,  nearly 20 times larger than the largest frequency reported by Yin \etal~\cite{Yin2021}. Finally, our field angle dependent data establish that these frequencies are associated with quasi-two-dimensional Fermi surface sheets.


\section{Methods}


\begin{figure}[!h]\centering
      \resizebox{8.5cm}{!}{
              \includegraphics{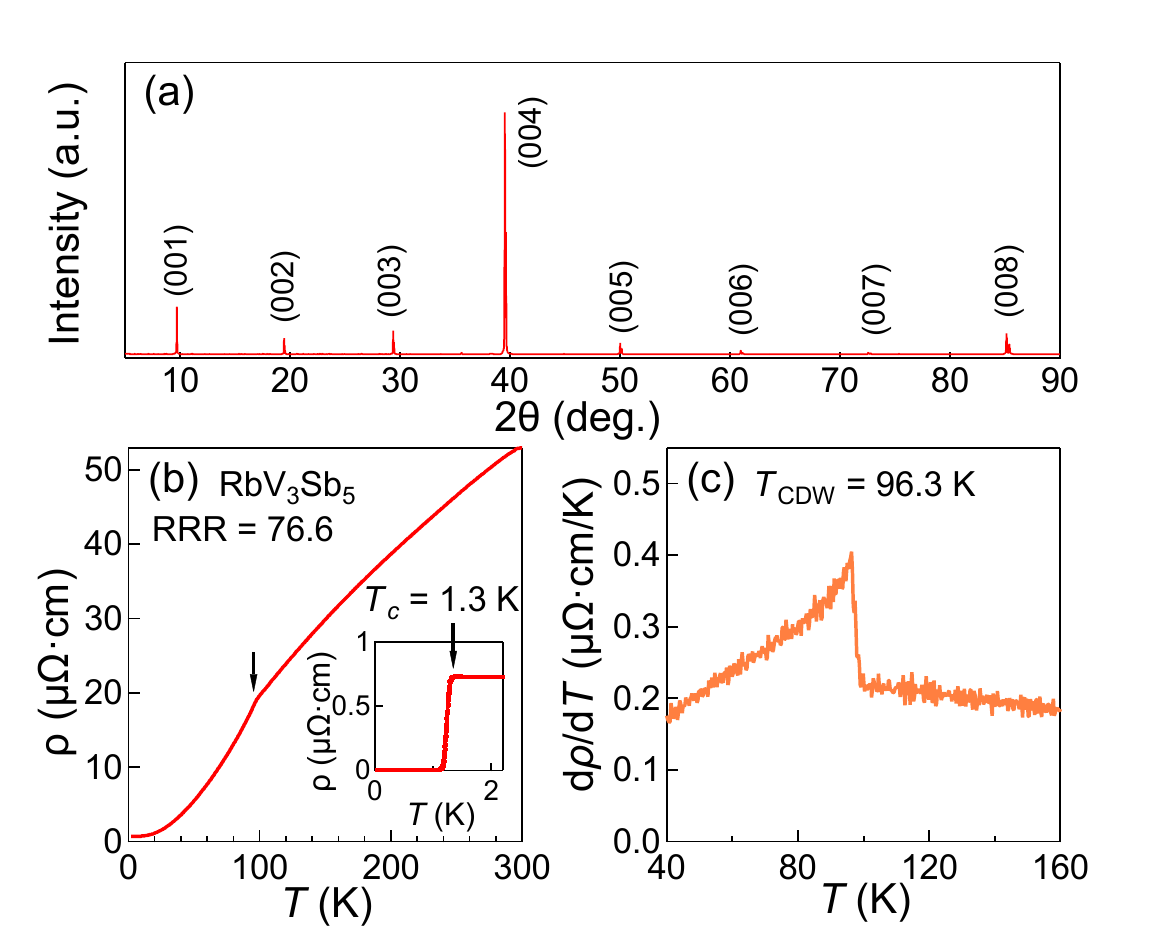}}                				
              \caption{\label{rhoT}(a) X-ray diffraction pattern of a \RbVSb\ single crystal showing only ($00L$) peaks. (b) Temperature dependence of the resistivity of \RbVSb\ under a zero external magnetic field. The RRR is 76.6. The inset displays the superconducting transition starting at 1.3~K. (c) Temperature dependence of d$\rho$/d$T$, displaying a sharp peak at $T_{\rm{CDW}}$ = 96.3~K. }
\end{figure}


Single crystals of \RbVSb\ were synthesized from Rb (ingot, 99.95 $\%$), V (powder, 99.9 $\%$) and Sb (shot, 99.9999 $\%$) using self-flux method similar to Ref.~[\onlinecite{Yin2021}].  Raw materials with the molar ratio of Rb:V:Sb = 5.5:7:14 were sealed in a quartz tube. To avoid the possible influence of the ambient environment, all preparation processes were performed in an argon-filled glovebox, before the quartz tube was moved to the furnace for heat treatment. The mixture was first heated to 1000$^\circ$C with the rate of 20$^\circ$C/h. After being held at 1000~$^\circ$C for 24 h, it was cooled to 900$^\circ$C at 50~$^\circ$C/h, followed by a further cooling to 400$^\circ$C at 2~$^\circ$C/h. The excess flux was removed by distilled water and ethanol. The as-grown single crystals were millimeter-sized shinny plates. 

X-ray diffraction (XRD) data were collected at room temperature by using a Rigaku X-ray diffractometer with Cu$K_\alpha$ radiation. The chemical compositions were characterized by a JEOL JSM-7800F scanning electron microscope equipped with an Oxford energy-dispersive X-ray EDX spectrometer. 

For magnetotransport measurements, the sample was firstly exfoliated from a bulk crystal and then transferred to a diamond surface pre-patterned with Au/Ti electrodes in a standard six-probe Hall bar configuration. A polydimethylsiloxane (PDMS) film were used to perform the transfer and Dupont 6838 silver paste was used to connect external leads with the electrodes. The transfer process was also performed in the glove box, and in order to protect the sample, we capped the sample with a thin layer ($\sim 500$~nm) of h-BN.  The thickness of the thin flakes and h-BN was determined by a dual-beam focused ion beam system (Scios~2 DualBeam by Thermo Scientific). These steps are similar to procedures optimized by us in Refs.~\onlinecite{Xie2021,Ku2022}.

Low-temperature and high-field measurements were carried out in a commercial Physical Property Measurement System (PPMS) by Quantum Design. A Stanford Research 830 lock-in amplifier was used for Shubnikov-de Haas measurements, while the standard PPMS Resistivity option was used for recording magnetoresistance and Hall effect. The rotator insert option by Quantum Design was used to tilt the angle between magnetic field and the $ab$ plane of the sample. During the rotation, the current direction is kept perpendicular to the magnetic field. A dilution fridge was additionally employed to measure the superconducting transition in RbV$_3$Sb$_5$.

 \begin{figure*}
    \centering
    \setlength{\abovecaptionskip}{0cm}
      \resizebox{16cm}{!}{
              \includegraphics[width=1\textwidth]{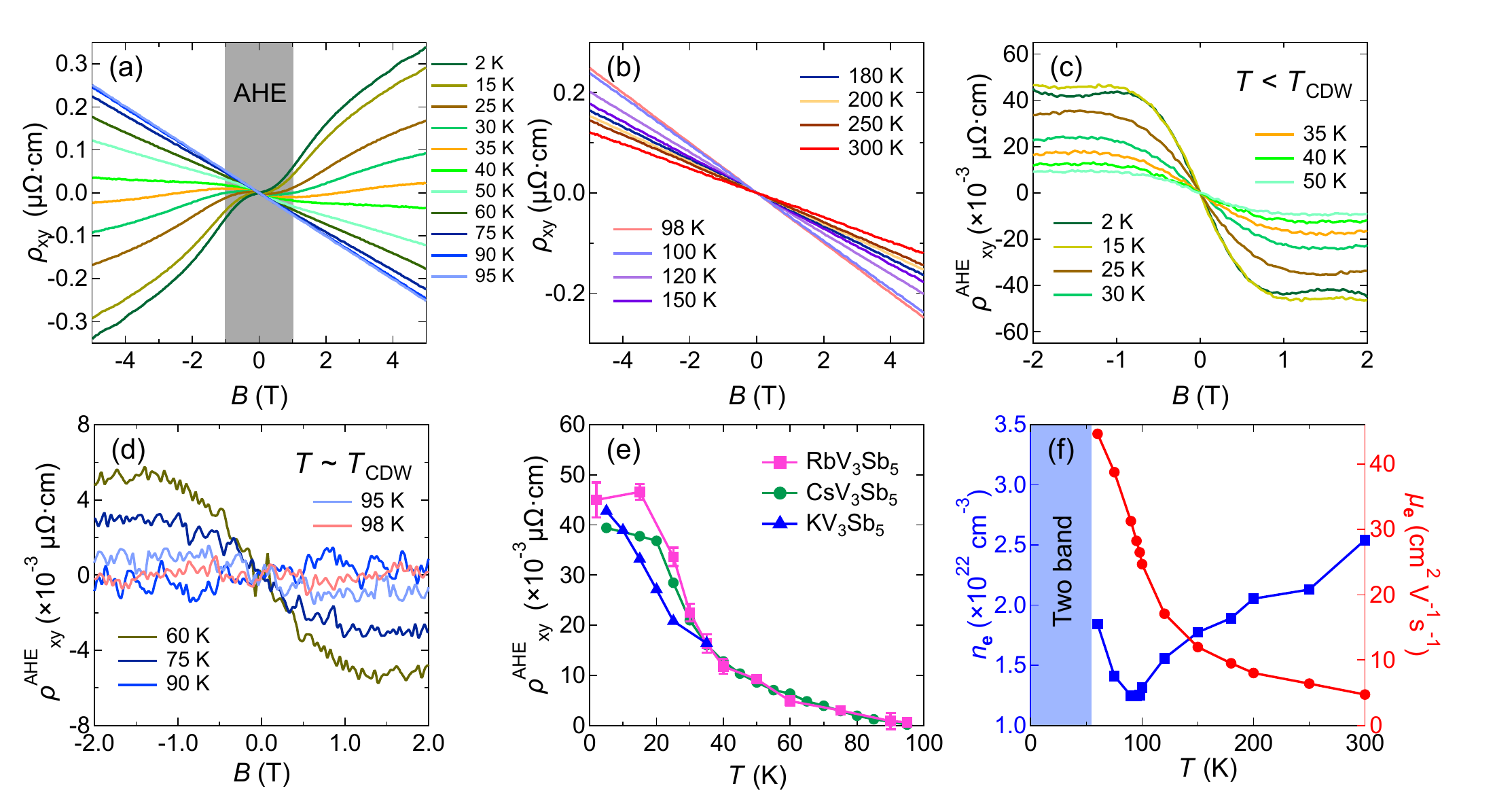}}                	
              \caption{\label{AHE}  
          (a), (b) Field dependence of Hall resistivity at various temperatures. The grey shading area represents the anomalous Hall effect (AHE) observed in the low-field region at all temperatures below $T_{\rm CDW}$. (c), (d) Anomalous Hall resistivity $\rho_{\rm xy}^{\rm AHE}$ at various temperatures after subtracting the ordinary Hall signal. The AHE shows up for all temperature below $T_{\rm CDW}$. (e) $\rho_{\rm xy}^{\rm AHE}$ in $\rm RbV_3Sb_5$, $\rm KV_3Sb_5$ and $\rm CsV_3Sb_5$ as a function of temperature. The data of $\rm KV_3Sb_5$ and $\rm CsV_3Sb_5$ are adapted from Refs.~[\onlinecite{Yang2020}] and [\onlinecite{Yu2021}], respectively. (f) Temperature dependence of  electron carrier density $n_e$ and mobility $\mu_e$ in $\rm RbV_3Sb_5$ extracted from the ordinary Hall signal in the one-band regime.
              }
\end{figure*}


\section{Results}

Figure~\ref{rhoT}(a) displays the XRD pattern of an as-grown single crystal of \RbVSb. The data can be indexed by the (00$L$) patterns from the crystal structure of \RbVSb~(space group: $P6/mmm$) reported in previous studies \cite{Ortiz2019,Yin2021}. This shows that our sample has a preferred orientation, as expected for a single crystal of layered materials. The chemical composition of the single crystal is obtained to be Rb:V:Sb = 0.94:2.86:5 via EDX (see the Supplementary Information), which further confirms the result of XRD in Fig.~\ref{rhoT}(a).

Figure~\ref{rhoT}(b) shows the temperature dependence of the electrical resistivity ($\rho(T)$) of our \RbVSb\ sample. The residual resistivity ratio (RRR) (defined as $\rho(300 {\rm K})/\rho(2 {\rm K})$) is about 77, currently one of the highest reported values. The d$\rho$/d$T$ vs. $T$ relation is shown in Fig.~\ref{rhoT}(c), where an anomaly can be clearly identified at 96.3~K. We associate this anomaly with $T_{\rm CDW}$, and our value is slightly lower than the reported value for this compound ~\cite{Wang2021,Cho2021,Yin2021}.

The availability of high-quality single crystals enables us to investigate the important transport phenomena in \RbVSb. We are particularly interested in anomalous Hall effect and quantum oscillations. We have performed Hall measurements over a wide range of temperatures. Figure~\ref{AHE}(a) shows the Hall response below $T_{\rm CDW}$ while Fig.~\ref{AHE}(b) shows that at temperatures higher than $T_{\rm CDW}$. The magnetic field ($B$) is parallel to the $c$-axis. At $T>T_{\rm CDW}$,  the Hall resistivity ($\rho_{xy}$) is perfectly linear in $B$, with the slope of $\rho_{\rm xy}$(B) curves gradually decreases upon cooling. The linearity persists up to $\pm$14~T, as shown in Fig.~S1. At temperatures near $T_{\rm CDW}$, $\rho_{\rm xy}(B)$ starts to deviate from the linear behavior and an antisymmetric, sideway ``S" shape begins to emerge in the low-field region, as highlighted by the grey shading region in Fig.~\ref{AHE}(a). With a further cooling towards the base temperature, the ``S"-shaped feature becomes even more pronounced. 
 
The Hall response of \RbVSb\ is reminiscent of the observation in sister compounds \KVSb\ and \CsVSb. In particular, the emergence of the low-field ``S"-shaped region can be attributed to the development of the anomalous Hall effect on the backdrop of the ordinary Hall effect. To extract the anomalous Hall component $\rho_{\rm xy}^{\rm AHE}$, we subtract a linear background from the total Hall signals using the data from $-2$~T to $+2$~T. The slope of the linear background is temperature-dependent, and it is derived from the locally linear region between 1~T and 2~T at all temperatures below $T_{\rm CDW}$. This analysis results in traces shown in Figs.~\ref{AHE}(c) and \ref{AHE}(d).
Using the $\rho_{\rm xy}^{\rm AHE}$ at the plateau region as the benchmark, we trace the temperature dependence of the AHE (see Fig.~\ref{AHE}(e)). It is worth noticing that the magnitude of $\rho_{\rm xy}^{\rm AHE}$ monotonically decreases with an increasing temperature, and becomes significantly weakened near $T_{\rm CDW}$. In fact, the disappearance of $\rho_{\rm xy}^{\rm AHE}$ coincides with $T_{\rm CDW}$, indicating an intimate relationship between the anomalous Hall effect and the CDW phase. Also included in Fig.~\ref{AHE}(e) are $\rho_{\rm xy}^{\rm AHE}$ of \CsVSb\ and \KVSb\ from Refs.~\onlinecite{Yu2021b} and \onlinecite{Yang2020}. Between $\sim$35~K and $\sim$95~K, $\rho_{\rm xy}^{\rm AHE}(T)$ of \RbVSb\ and \CsVSb\ overlap almost perfectly. Furthermore, these data connect smoothly to $\rho_{\rm xy}^{\rm AHE}$ of \KVSb\ near 35~K. Finally, we extract the mobility ($\mu_e$) and the carrier density ($n_e$) for $\rho_{\rm xy}(B)$ traces that are linear over a wide field range, using a simple single band model. We find that $n_e$ exhibits a minimum right at $T_{\rm CDW}$ while $\mu_e$ is significantly enhanced upon cooling across $T_{\rm CDW}$. These temperature dependence are similar to the observation reported in \KVSb.


\begin{figure*}
    \centering
      \resizebox{16cm}{!}{
              \includegraphics[width=1\textwidth]{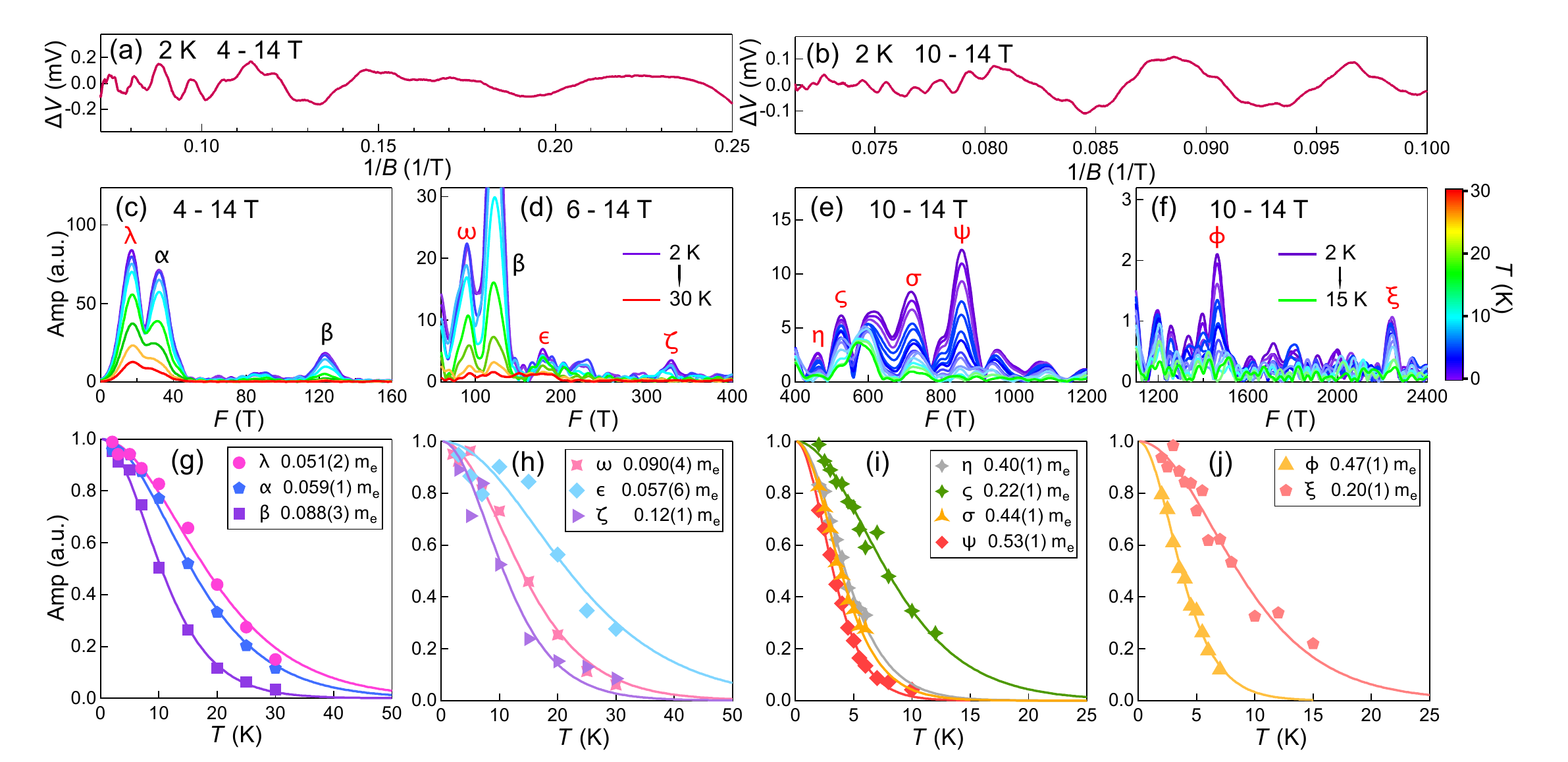}}                
              \caption{\label{qo}  (a, b) Oscillation signals after removing the background. (c-f) Fast Fourier transform (FFT) spectra in four different ranges: (c) 0-160 T, (d) 60-400 T, (e) 400-1200 T, (f) 1100-2400 T. The field ranges used for FFT analyses are 4-14~T, 6-14~T, 10-14~T and 10-14~T in (c), (d), (e) and (f), respectively. The data were collected at various temperatures ranging from 2~K to 30~K. The peaks are marked in the figures as $\lambda \sim17$~T, $\alpha\sim34$~T, $\beta\sim121$~T, $\omega\sim91$~T, $\epsilon\sim178$~T, $\zeta\sim322$~T, $\eta\sim461$~T, $\varsigma\sim525$~T, $\sigma\sim716$~T, $\psi\sim856$~T, $\phi\sim1461$~T and $\xi\sim2235$~T. The peaks labelled in red are newly discovered in our findings, while those labelled in black have been reported previously. (g-j) Temperature dependence of FFT amplitude of the peaks obtained in (c-f) and the corresponding  Lifshitz-Kosevich fittings. The estimated effective masses are shown in the figures. }
\end{figure*}


We now present the second element obtained from our magnetotransport data. In Figs.~\ref{qo}(a) and \ref{qo}(b), we show the SdH oscillations collected with $B\parallelsum c$ after removing the background. Two different field ranges are shown: a long field range offers a better view of low-frequency oscillations, while a narrower field range is adopted to display the finer structure in our data due to higher SdH frequencies.
Fast Fourier transform (FFT) spectra are shown in Figs.~\ref{qo}(c-f). We detect a rich structure, uncovering at least 12 SdH peaks in the range of between $0$ and $3987$~T, where 3987~T is the area of the reconstructed Brillouin zone in the $k_x$-$k_y$ plane expressed as a quantum oscillation frequency.
$F_\alpha$ and $F_\beta$ are consistent with the previous report by Yin \etal~\cite{Yin2021}, but the remaining peaks have never been reported.

In Figs.~\ref{qo}(c-f), the new peaks are marked with red Greek symbols. Following the convention proposed for \CsVSb,~\cite{Zhang2022a} we regard the SdH frequencies lower than 500~T as belonging to the `low-frequency spectrum'. In the literature of \CsVSb, low-frequency spectrum has been presented in almost all reports, with excellent consistency. On the other hand, peaks in the mid-frequency spectrum (between 500~T and 2000~T) and the high-frequency spectrum (greater than 2000~T) of \CsVSb\ are absent in many report. In \RbVSb, our $F_\alpha$ and $F_\beta$ are similarly in good agreement with the previous report~\cite{Yin2021}. However, our data additionally reveal peaks in both the mid-frequency to the high-frequency spectra (Figs.~\ref{qo}(e) and (f)). 
The largest frequency ($F_\xi$) reaches 2235~T. These frequencies indicate that \RbVSb\ in the CDW phase hosts multiple Fermi surface sheets, and the largest cyclotron orbit encloses an area that is $\sim$56\% of the Brillouin zone area.

 Following the Lifshitz-Kosevich theory, we analyze the temperature dependence of the oscillation amplitudes using the thermal damping factor $R_T=X/\sinh X$ with $X = 14.693m^* T/B$, from which we can calculate the cyclotron effective mass $m^*$ of the peaks marked by the Greek symbols. As shown in Figs.~\ref{qo}(g-j), $m^*$ of the low-frequency peaks ($\lambda $, $\alpha$, $\beta$, $\omega$, $\epsilon$ and $\zeta$) are relatively small, all less than 0.1$m_e$ except for $\zeta$, which is slightly higher (0.12$m_e$). However, $m^*$ of the mid-frequency and the high-frequency peaks are comparatively larger, all greater than 0.2$m^*$ with $\psi$ having the largest value of 0.53$m_e$. Finally, we note our observation that some peaks shift at higher temperatures. We attribute this phenomenon to noise contamination. Indeed, this happens to weak peaks that are close to the noise floor, and they are excluded from our analysis. For instance, only the data between 2~K and 6~K are included for $m^*$  analysis for $\eta$ and $\sigma$ (see Figs.~\ref{qo}(e) and ~\ref{qo}(i)). 

 \begin{figure}[!t]\centering
      \resizebox{8.5cm}{!}{
             \includegraphics{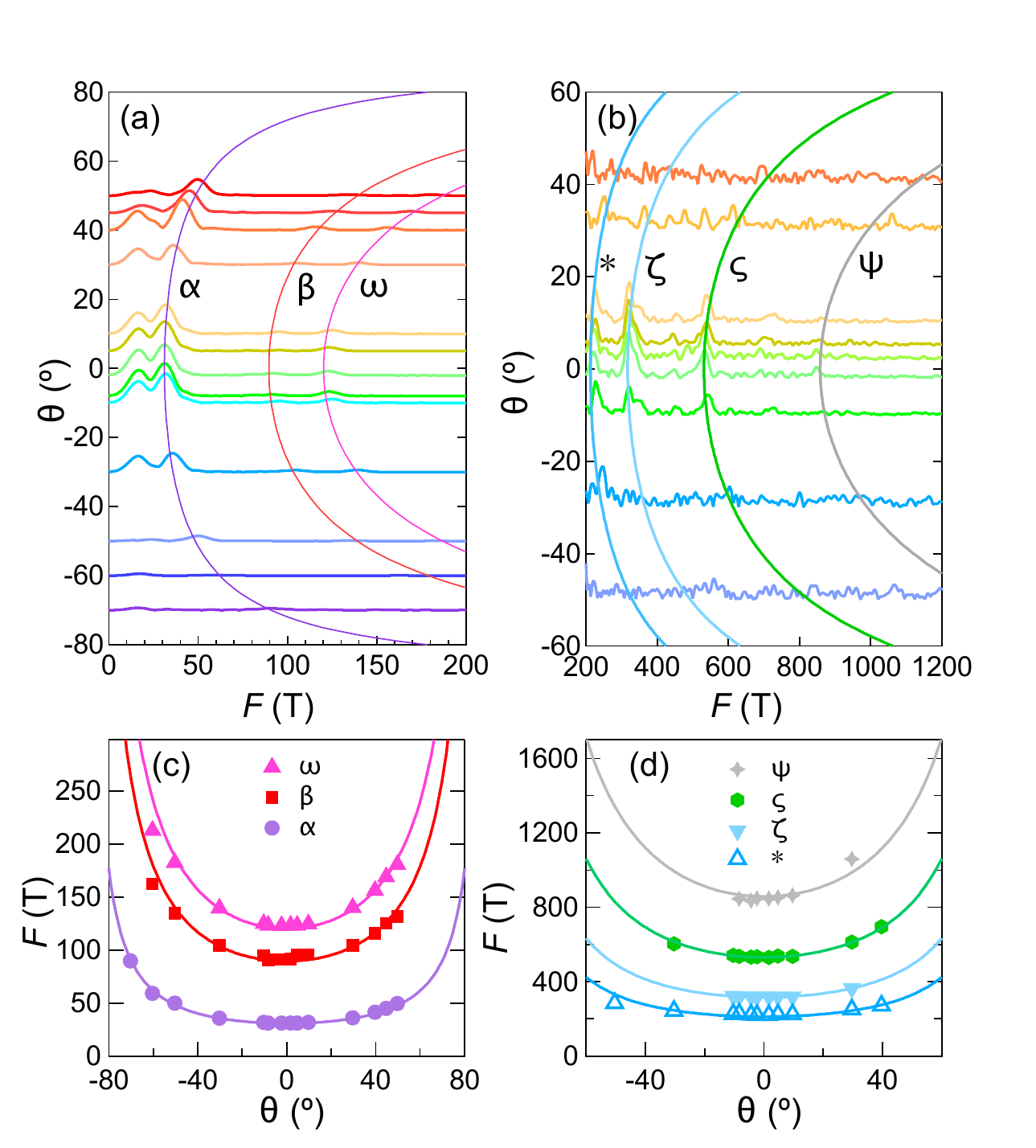}}      
             \caption{\label{ang_dep} 
              FFT spectra of SdH oscillations at different field angles ($\theta$) for frequencies ranging from (a) 0 to 200~T and (b) 200~T to 1200~T. All peaks shift towards higher frequencies as $\theta$ increases.  (c, d) Angular dependence of different oscillation frequencies. The curves in all panels are the fittings based on the formula $F(\theta) = F(0^{\circ})/\cos\theta$. }

\end{figure}


We now explore the dimensionality of the Fermi surface. Figure~\ref{ang_dep} shows the FFT spectra at 2~K for selected magnetic field angles $\theta$, where $\theta=0^\circ$ corresponds to $B\parallelsum c$ and $B$ is rotated towards the sample $ab$-plane with an increasing $\theta$.  The intensities of all peaks gradually decrease with an increasing angle, especially for the peaks with relatively high frequencies ($>$ 500~T). However, we are able to follow several stronger peaks to higher angles.
$F_\alpha$, $F_\beta$ and $F_\omega$, which are the more intense peaks, can be traced with ease to $\sim70^\circ$. For higher frequencies $F_\varsigma$, $F_\psi$ and $F_\zeta$, the signals are weaker but we are able to follow their angular evolution to $\sim35^\circ$. All these frequencies display $1/\cos\theta$ dependence (Fig.~\ref{ang_dep}), indicating the quasi-two-dimensional nature of \RbVSb. Additionally, we recorded a peak with $F(0^\circ)=212$~T, which also exhibits a $1/\cos\theta$ dependence (see the curves labelled by asterisk in Figs.~\ref{ang_dep}(b) and (d)). However, the temperature dependence of this peak does not conform to the description of $R_T$.


\begin{figure}[!t]\centering
       \resizebox{8.5cm}{!}{
              \includegraphics{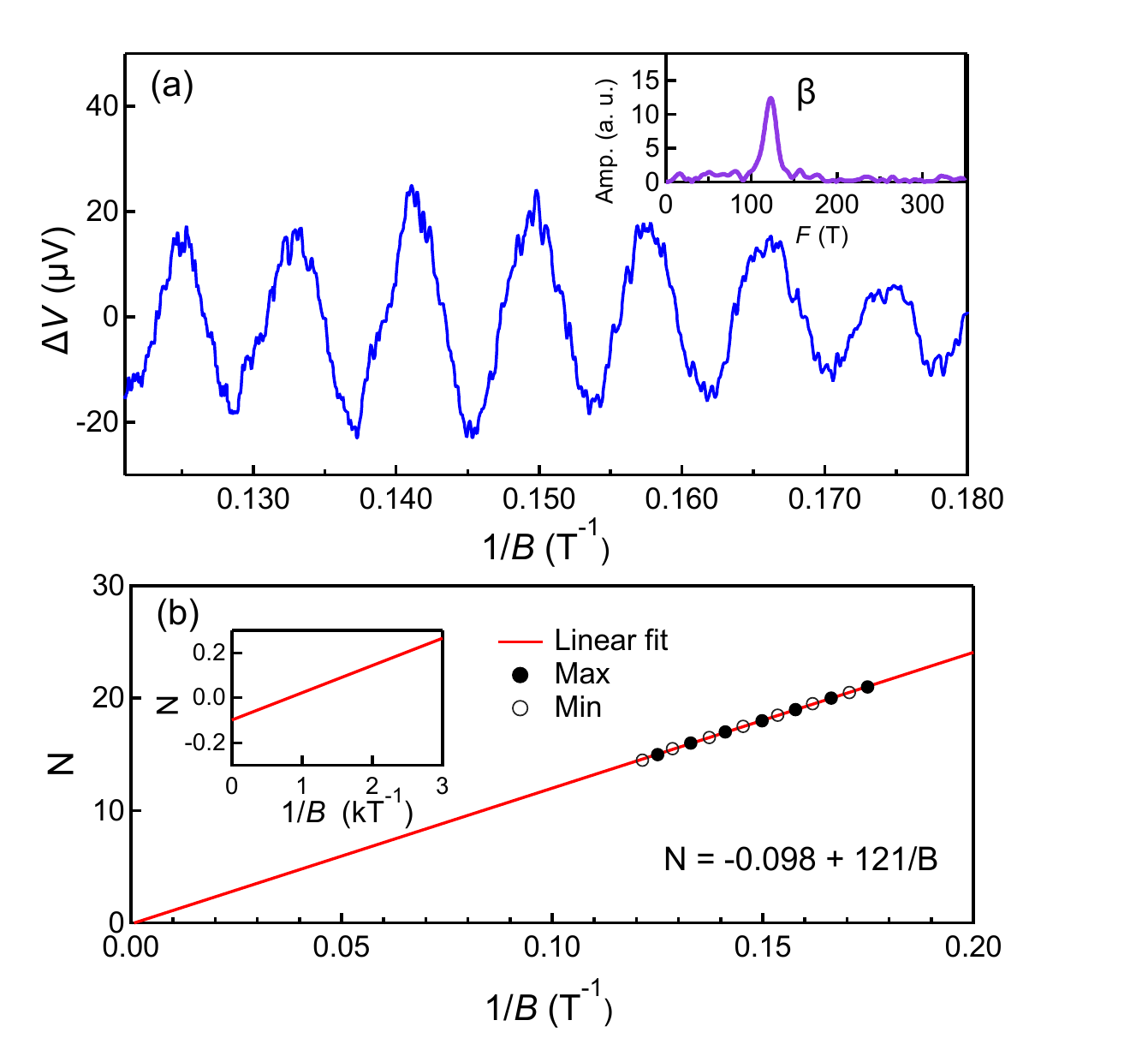}}                			
              \caption{\label{LL_fan}
              (a) The oscillatory signal $\Delta V$ (proportional to oscillatory $\rho_{xx}$) of  $F_\beta$ as a function of inverse magnetic field $B^{-1}$. A high pass digital filter was used to eliminate the intense $\lambda$ and $\alpha$ signals. The inset shows the FFT spectrum of the filtered signal. (b) Landau level fan diagram analysis of $F_\beta$ constructed with the filtered signal shown in (a). The closed circles denote integer indices while the open ones indicate half-integer indices. The red line is a linear fit of the data. The inset shows an enlarged view near the origin.}

\end{figure}

The total oscillatory signals can be decomposed into constituent components, with the $i^{th}$ component taking the form of~\cite{Jin2016,Fu2021,Shoenberg1984} 

\begin{equation}
 \Delta \rho_{xx}^i \sim A_i\cos\left(2\pi\left(\frac{F_i}{B}+\gamma_i-\delta_i \right)\right),
 \label{LK}
\end{equation}
\\ 
where $A_i$ is the oscillation amplitude that depends on the relevant damping factors. For the phase factor,  $\gamma_i=1/2-{\phi_B}_i/2\pi$ where ${\phi_B}_i$ is the Berry phase\cite{Mikitik1999}. $\delta_i$ depends on the dimensionality of the Fermi surface, and it is equal to $0$ or $\pm \frac{1}{8}$ when the Fermi surface is two-dimensional or three-dimensional, respectively\cite{Xiang2015}. We construct a Landau level (LL) fan diagram for $\beta$  based on the data at 2~K, as $\beta$ is the easiest component to isolate with a digital high-pass filter. The high-pass filter removed the contribution from $\lambda$ and $\alpha$, making $\beta$ the dominant component (the components with $F>F_\beta$ makes negligible contribution.) Figure \ref{LL_fan}(a) shows the filtered signal and the inset is the corresponding FFT spectrum, showing the dominance of $\beta$ band. At $\pm14$~T, $\rho_{\rm xx}/\left|\rho_{\rm xy}\right|\sim5$ (shown in Fig.~S1) and this ratio significantly increases with a decreasing external field. As a result, $\sigma_{\rm xx} = \frac{\rho_{\rm xx}}{\rho_{\rm xx}^2+\rho_{\rm xy}^2} \approx \frac{1}{\rho_{\rm xx}}$, showing that $\rho_{\rm xx}$ reaches a local minimum when $\sigma_{\rm xx}$ reaches a local maximum  and vice versa. Our oscillatory voltage $\Delta V$ is proportional to oscillatory resistivity $\Delta\rho_{xx}$. Thus, the LL integer indices $N$ can be safely assigned to the local maxima of the signal while the LL half-integer indices $N+1/2$ are assigned to the minima , as shown in Fig.~\ref{LL_fan}(b).  We then perform the linear fitting of $N$ as a function of $1/B$. The fitted slope is 121~T, identical to the frequency of $\beta$, suggesting the reliability of our analysis. The intercept on the LL index axis is $-0.098$. Considering the two-dimensional nature of $\beta$ as shown in Fig.~\ref{ang_dep}(a), the value of $\delta_\beta$ should be zero. Finally we can obtain $\phi_B = 2\pi\times (-0.098)=-0.196 \pi$ . This value is close to 0, indicating a trivial Berry phase for $\beta$. This result shows that $\beta$ band does not contribute to $\sigma_{\rm xx}^{\rm AHE}$ intrinsically, while the topology of other bands should attract further study.

 \section{Discussion}
The fact that \RbVSb\ is also able to support the giant anomalous Hall effect brings new insights to the study of \AVSb\ compounds. One remarkable feature deserving further investigation is the good agreement across all three \AVSb\ compounds (see Fig.~\ref{AHE}(e)), considering the fact that these measurements were collected on different crystals by independent groups. This suggests similar origins of AHE. Besides, a similar scaling relation between $\sigma_{\rm xy}^{\rm AHE}$ and $\sigma_{\rm xx}$ observed in \RbVSb\ and the other two compounds also supports this opinion. As plotted in Fig.~S2, we can identify a region where $\sigma_{\rm xy}^{\rm AHE}$ $\propto \sigma_{\rm xx}^2$ when $\sigma_{\rm xx}$ exceeds $\sim2\times10^{5} \rm\,  \Omega^{-1}cm^{-1}$.  

Judging from the magnitude of $\sigma_{\rm xx}$, skew scattering has been put forward to explain the observed AHE in \CsVSb\ and \KVSb\ (Refs.~\onlinecite{Nagaosa2010,Yang2020,Yu2021b}). However, the mechanism giving rise to skew scattering is still under debate. In \KVSb, it is attributed to a spin cluster model with an extra spin-orbit coupling. In \CsVSb, tiny amounts of paramagnetic impurities could cause the skew scattering. However, as pointed out by Yu \etal~\cite{Yu2021b}, it is not clear if a very low level of paramagnetic impurities is sufficient to cause the large AHE observed. When it comes to \RbVSb, skew scattering is likely to be a dominant mechanism, as the region of $\sigma_{\rm xx}$ where the large AHE is observed is similar to \KVSb\ and \CsVSb. However, the source of the skew scattering requires more studies. Nevertheless, the possibility of an intrinsic mechanism giving rise to the AHE should still be investigated, as TRSB has been observed in the CDW state without the external magnetic field. 
 All the evidence indicates a topological band structure could be a potential source of intrinsic anomalous Hall effect in \AVSb. Unfortunately, our LL fan diagram analysis only covers $\beta$ frequency. Further study on other frequencies is urgently needed. 

The observation of the AHE in the CDW state in all three \AVSb\ compounds clearly highlights the unconventional nature of this phase, demonstrating the need to understand the CDW state. In this regard, the Fermi surface of these compounds in the CDW state is an important gateway to achieve this understanding. Density functional theory (DFT) calculations performed on \CsVSb\ assuming either a `star-of-David' (SoD) or a tri-hexagonal (TrH) deformation, the so-called $2\times2\times1$ distortion, can not explain our rich spectrum in \RbVSb. It has been pointed that additional interlayer complexity that involves both SoD and TrH can introduce new quantum oscillation frequencies~\cite{Broyles2022, Zhang2022a}. Therefore, future DFT calculations on \RbVSb\ taking into account various stacking arrangement of SD and ISD will shed light on this issue, and our quantum oscillation data will potentially be useful for settling the superlattice structure in the CDW phase of \RbVSb.

\section{Conclusions}

To summarize, we have synthesised high-quality single crystals of \RbVSb\ and conducted the magnetotransport measurements up to 14~T. Close to the zero field, we detect the emergence of the anomalous Hall effect, while at high fields, we unravel a rich quantum oscillation spectrum. The magnitude of the anomalous Hall resistivity is large, on par with the observation in the sister compounds \CsVSb~\cite{Yu2021b} and \KVSb~\cite{Yang2020}, highlighting \AVSb\ as a suitable platform to explore the physics of anomalous Hall resistivity. Regarding quantum oscillations, we report 12 frequencies, with the largest frequency having a value of 2235~T. Angular dependencies of some of these frequencies show that they originate from two-dimensional Fermi surfaces, consistent with the layered nature of \RbVSb.

\begin{acknowledgments}
This work was supported by Research Grants Council of Hong Kong (CUHK 14301020, CUHK 14300722, A-CUHK402/19), CUHK Direct Grant (4053461, 4053408, 4053528, 4053463), the National Natural Science Foundation of China (12104384, 12174175) and the Shenzhen Basic Research Fund (JCYJ20190809173213150).\\ 
$^\S$L.W., W.Z. and Z.W. contributed equally to this work.
\end{acknowledgments}

\providecommand{\noopsort}[1]{}\providecommand{\singleletter}[1]{#1}%

\end{document}